\documentclass[useAMS,usenatbib,12pt,psfig]{mn2e} 
\usepackage{color}
\usepackage{graphicx}
\usepackage{txfonts}

\def\mh1{$M_{\rm H_{I}}$}

\title[H{\sc i} in Arp72 and M51-type systems]{H{\sc i} in Arp72 and similarities with M51-type systems}
\author [Sengupta {\it{et al.}}]{ Chandreyee Sengupta,$^{1}$\thanks{e-mail:sengupta@ncra.tifr.res.in(CS)
  djs@ncra.tifr.res.in(DJS),  dwaraka@rri.res.in(KSD)} D.J. Saikia$^{2}$ and K. S. Dwarakanath$^{3}$  \\\\
$^{1}$ Calar Alto Observatory, IAA$-$CSIC, Spain \\
$^{2}$ National Centre for Radio Astrophysics, Tata Institute of Fundamental Research, Pune 411 007, India \\
$^{3}$ Raman Research Institute, Bangalore 560 080, India}

\begin{document}

\date{Received  ; accepted  }
\date{}
\pagerange{\pageref{firstpage}--\pageref{lastpage}} \pubyear{}

\maketitle

\label{firstpage}

\begin{abstract}
We present neutral hydrogen (H{\sc i}) observations with the Giant Metrewave Radio Telescope ({\it GMRT}) of the 
interacting galaxies NGC5996 and NGC5994, which make up the Arp72 system. Arp72 is an M51-type system and shows
a complex distribution of H{\sc i} tails and a bridge due to tidal interactions. H{\sc i} column densities 
ranging from 0.8$-$1.8$\times$10$^{20}$ atoms cm$^{-2}$ in the eastern tidal tail to 1.7$-$2$\times$10$^{21}$ 
atoms cm$^{-2}$ in the bridge connecting the two galaxies, are seen to be associated with star-forming regions.  
We discuss the morphological and kinematic similarities of Arp72 with M51, the archetypal 
example of the M51-type systems, and Arp86, another M51-type system studied with the {\it GMRT}, and suggest that  
a multiple passage model of Salo \& Laurikainen may be preferred over the classical single passage model 
of Toomre \& Toomre, to reproduce  the H{\sc i} features in Arp72 as well as in other M-51 systems depicting 
similar optical and H{\sc i} features.
\end{abstract}


\begin{table*} 
\caption{GMRT observations }
\begin{tabular}{l l l l l l l rrr}
\hline
Frequency & Observation  & Phase      & Phase cal    &  $\tau$ & Bandwidth &rms (per channel  & \multicolumn{3}{c}{Beam size} \\ 
          & date         & calibrator &  flux        &         &           &for 21-cm line)   &  maj & min & PA                \\
          &              &            &density (Jy)  & (hr)    & (MHz)     &(mJy beam$^{-1}$) &($^{\prime\prime}$)&($^{\prime\prime}$)
&($^\circ$)\\ 
\hline
 21-cm line & 2008 Apr 30 &J1609+266 &  5.0 & 7   & 8 &  0.5 &   9.0 &  9.0 &      \\
            &             &          &      &     &   &  0.9 &  29.0 & 21.0 & 119  \\
            &             &          &      &     &   &  1.0 &  41.0 & 35.0 & 126   \\
  1403 MHz  &             &          &      &     &   &  0.17&   3.5 & 2.2 & 40  \\
\hline

\end{tabular}
\end{table*}

\section{Introduction}

The role of tidal interactions, collisions and mergers of galaxies in driving and
shaping galaxy evolution over cosmic time scales, 
often triggering bursts of star formation, has been widely recognized
since the early work by Toomre \& Toomre (1972) as well as from many
recent pieces of work (see Kennicutt, Schweizer \& Barnes 1998; Barnes \& Hernquist 1992; 
Barnes 1999; Struck 1999; Schweizer 2000, 2005).
However, studies of gravitational interactions
at optical wavelengths have often focussed on major mergers which involve galaxies of
similar mass, leading to spectacular structures. Minor
mergers with mass ratios in the range of $\sim$0.1$-$0.5 may appear less spectacular at optical
wavelengths, but show clear and complex signs of
interactions in their H{\sc i} distributions as seen for example in M51 (Rots et al. 1990)
and Arp86 (Sengupta, Dwarakanath \& Saikia 2009). Although such minor mergers and interactions have
so far usually been studied in the nearby Universe, more sensitive H{\sc i} observations with
present and upcoming telescopes should enable us to examine such interactions at higher redshifts
and add to our knowledge of galaxy growth and evolution through minor mergers and interactions.
An important parameter to quantitatively assess the importance of interactions and mergers is
the merger rate, which is the number of mergers per unit time and unit co-moving volume, which
appears to increase with redshift (e.g. Abraham et al. 1996; Ryan et al. 2008). Identification, 
imaging and modelling of a large number of M51-type systems will give us a better idea of the 
range of merger time-scales and hence the merger rate for these minor mergers and interactions.

In addition to mergers and interactions, there has also been a lot of 
interest in understanding how star formation rates are affected by collisions and mergers
of galaxies since the early pioneering pieces of work by Larson \& Tinsley (1978) and
Struck-Marcell \& Tinsley (1978). Observations with the Infrared Astronomy Satellite
{\it (IRAS)} drew attention to the highly luminous infrared galaxies, the more extreme
ones with far infrared luminosities greater than 10$^{12}$L$_\odot$ being christened as
ultraluminous infrared galaxies \citep{Sanders88}. These are believed to be the result of mergers of 
similar-mass, gas-rich galaxies (see Soifer et al. 1987; Smith et al. 1987; Sanders \& Mirabel
1996; Struck 1999). 
The high resolution of {\it Spitzer} infrared observations have made it possible to  study
individual star-forming regions (see Smith et al. 2007, and references therein). The advent
of the Galaxy Evolution Explorer {\it (GALEX)} ultraviolet (UV) telescope, has provided another
window to study stellar populations and star-formation history in galaxies (see Smith et al.
2010, and references therein). Ultraviolet observations show that tidal features are 
often quite bright at these wavelengths (e.g. Neff et al. 2005;  Hancock et al. 2007), 
and are sometimes also known to reveal new tidal features, as seen in 
the interacting galaxy NGC4438 in the Virgo cluster
(e.g. Hummel \& Saikia 1991; Hota, Saikia \& Irwin 2007) by Boselli et al. (2005). At radio
frequencies, both H{\sc i} and CO observations have been good tracers of gas distributions
for understanding the star-formation process (e.g. Helfer et al. 2003; Greve et al. 2005;
Leroy et al. 2008, 2009; Walter et al. 2008; Bigiel et al. 2008), while continuum observations
have been useful to constrain star-formation and supernova rates (see Condon 1992 for a review),
as well as probe deep into the circumnuclear regions to study the supernova remnants and 
H{\sc ii} regions, as in the archetypal starburst galaxies M82 (Fenech et al. 2008, and references
therein) and NGC1808 (Saikia et al. 1990; Collison et al. 1994). 

In this paper, we present the results of H{\sc i} observations of the interacting 
galaxies in Arp 72, which were observed as part of an ongoing H{\sc i} survey of interacting 
galaxies to study their H{\sc i} properties and the correlation of gas and stars in these 
galaxy pairs.  Arp72 is an M51-type system, and such systems have often been used as 
laboratories for understanding the possible dynamics that give rise to the striking spiral 
arms, and tidal bridges and tails, as well as the correlation of gas and stars in these features 
\citep{Salo, Salo2000, Reshetnikov}.  We also present a comparative analysis of our  results on
the two M51-type systems we have studied so far, Arp72 and Arp86, with M51. The results for Arp86 
were presented earlier by Sengupta, Dwarakanath \& Saikia (2009). For comparison with M51, 
we have used the results from The H{\sc i} Nearby Galaxy Survey (THINGS; Walter et al. 2008).  

\section{Arp72}
Arp72 is a nearby unequal-mass interacting galaxy pair undergoing an intense 
phase of star formation. The system consists of NGC5996, the bigger galaxy of morphological 
classification SBc, at a radial velocity of  3297$\pm$2 km~s$^{-1}$, and the smaller companion 
NGC5994, a possible barred spiral, at a radial velocity of 3290$\pm$10 km~s$^{-1}$. The angular 
diameters of NGC5996 and NGC5994 are $\sim$1.6$^{\prime}$ and 0.6$^{\prime}$ respectively, which 
translates to a physical extent of $\sim$20 and 8 kpc respectively, at a distance of 43.9 Mpc to the system
(1 arcsec = 0.21 kpc). 
Throughout this paper, 43.9 Mpc has been used as the distance to the system, which has been estimated 
using their average optical velocity and a Hubble constant of 75 km~s$^{-1}$~Mpc$ ^{-1}$. The 
morphological classifications, angular diameters and the optical radial velocities are from 
the NASA/IPAC Extragalactic 
Database (NED). NGC5996, is known to be a Markarian galaxy (Mrk691), with enhanced star formation 
possibly initiated by the interaction with NGC5994. The system has been a part of several statistical 
studies \citep{Kandalyan, schwartz-martin,woods}, but has not been studied systematically at radio 
frequencies. Recently Smith et al. (2010) have presented a {\it GALEX} image of the system which
shows a long `beaded' tail towards the east of the main galaxy, which although seen in the Sloan
Digital Sky Survey (SDSS) and Arp Atlas images, is more pronounced at UV wavelengths. 
The bridge connecting the two galaxies is also bright in the {\it GALEX} image. 
The bridge and the northern spiral arm are quite bright in H$\alpha$ (Smith et al. 2010),
and the oxygen abundance in the centre of the bridge, log[O/H]+12, is $\sim$8.7 (Hancock et al. 
in preparation; Smith et al. 2010).

\section {Observations}
Arp72 was observed for 7 hours in H{\sc i} 21-cm line on 2008 April 30, using the Giant 
Metrewave Radio Telescope {\it (GMRT)}. 
The full width at half maximum of the primary beam of {\it GMRT} antennas is $\sim$24\arcmin ~at 1420 MHz. 
The baseband bandwidth used was 8 MHz for the 21-cm H{\sc i} line observations giving a
velocity resolution of $\sim$13.7 km~s$^{-1}$. Some of the parameters of the observations and also the
rms noise and beam sizes for the results presented here are summarized in Table 1.

Data obtained with the {\it GMRT} were reduced using {\tt AIPS} (Astronomical Image Processing System). 
Bad data due to dead antennas and those with significantly lower gain than others, and radio frequency interference (RFI) 
were flagged and the data were calibrated for amplitude and phase using the primary and secondary calibrators. 
The primary flux density calibrator was 3C286, while the phase calibrator was J1609+266 (see Table 1).
The flux densities are on the scale of Baars et al. (1977) and the error on the flux density is $\sim$5 per cent.
The calibrated data were used to make both the H{\sc i} line images and the 1403-MHz radio continuum images 
by averaging the line-free channels and self calibrating. 
For the H{\sc i} line images the calibrated data were continuum subtracted using the AIPS tasks `UVSUB' and `UVLIN'. 
The task `IMAGR' was then used to get the final 3-dimensional deconvolved H{\sc i} data cubes. From these cubes the 
total H{\sc i} images and the H{\sc i} velocity fields were extracted using the AIPS task `MOMNT'.
To examine the structures on different levels in both H{\sc i} and
radio continuum, we produced images of different resolutions by tapering the data to different uv limits.

\section {Observational Results} 
\subsection {H{\sc i} morphology of Arp72}
The low-resolution H{\sc i} column density image of Arp72, with an angular resolution of 
$\sim$40$^{\prime\prime}$ which corresponds to a linear resolution of $\sim$8.5 kpc, overlaid on the
optical  r-band SDSS image is shown in the upper panel of Fig. 1. The 
corresponding H{\sc i} spectrum of the system is shown in the lower panel of Fig. 1. 
The integrated flux density estimated from this spectrum is 30.1 Jy~km~s$^{-1}$  and the corresponding 
H{\sc i} mass is 1.4$\times$10$^{10}$ M$_{\odot}$, for a distance of 43.9 Mpc to the system. 
Our total flux density is somewhat higher than the value of 22.97$\pm$0.34 Jy~km~s$^{-1}$ estimated
by Giovanardi \& Salpeter (1985) from an Arecibo spectrum of the galaxy pair (UGC10033) over a size of
3.8$\times$2.8 arcmin$^2$, and the HIPASS (H{\sc i} Parkes All Sky Survey; Barnes et al. 2001) 
estimate of $\sim$19.7 Jy~km~s$^{-1}$ \citep{Wong09}.
 The lower flux densities for the Arecibo and HIPASS observations are likely to be due
to the higher rms  noise values of these observations, which are $\sim$3 and 10 times higher than 
that of the {\it GMRT} observations. To explore this possibility, we convolved the low-resolution
{\it GMRT} cube ($\sim$40$^{\prime\prime}$) to be of similar resolution to the Arecibo beam ($\sim$3 arcmin)
and estimated the flux density using a threshold similar to the rms noise of the Arecibo
survey of $\sim$3 mJy. The integrated flux density decreased from our estimate of 30.1 Jy~km~s$^{-1}$
to 20.3 Jy~km~s$^{-1}$, indicating that higher rms noise values for observations of this source with
extended diffuse emission could lead to lower estimates of the integrated flux density.

Extended tidal tails and debris, expected from the interaction between NGC5996 and NGC5994, are seen 
around the system. An  H{\sc i} bridge, aligned to the optical and UV ones which connects
NGC5996 to its low-mass companion, is clearly seen. In addition, there is a long ($\sim$3.8 arcmin, $\sim$50 kpc) 
tidal tail that originates at the northern edge of the disk of the larger galaxy NGC5996, curves towards the east 
and then towards the south, while the smaller galaxy NGC5994 exhibits an  H{\sc i} extension towards the west. 
These features are reminiscent of those seen in M51 (Rots et al. 1990) and Arp86 (Sengupta et al. 2009).
The corresponding velocity field is shown in Fig. 2. 
The disk of NGC5996 shows regular rotation; however, signatures of tidal interaction with NGC5994 
are seen at the edge of the disk in the form of disturbance in that regular rotation pattern. 

\begin{figure}
     \hbox{
    \centerline{\rotatebox{0}{\includegraphics*[height=3.0in]{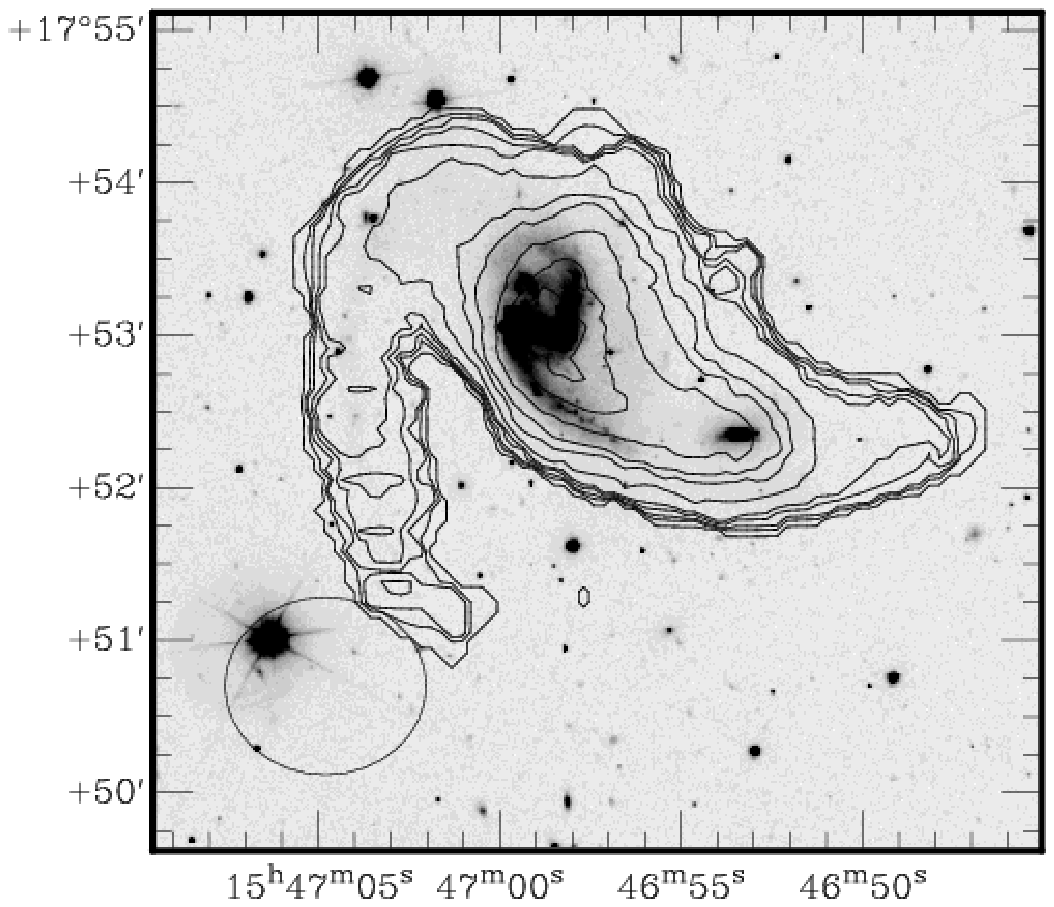}} }
    \hspace{-8.2cm}
    \centerline{\rotatebox{-90}{\includegraphics*[height=3.0in]{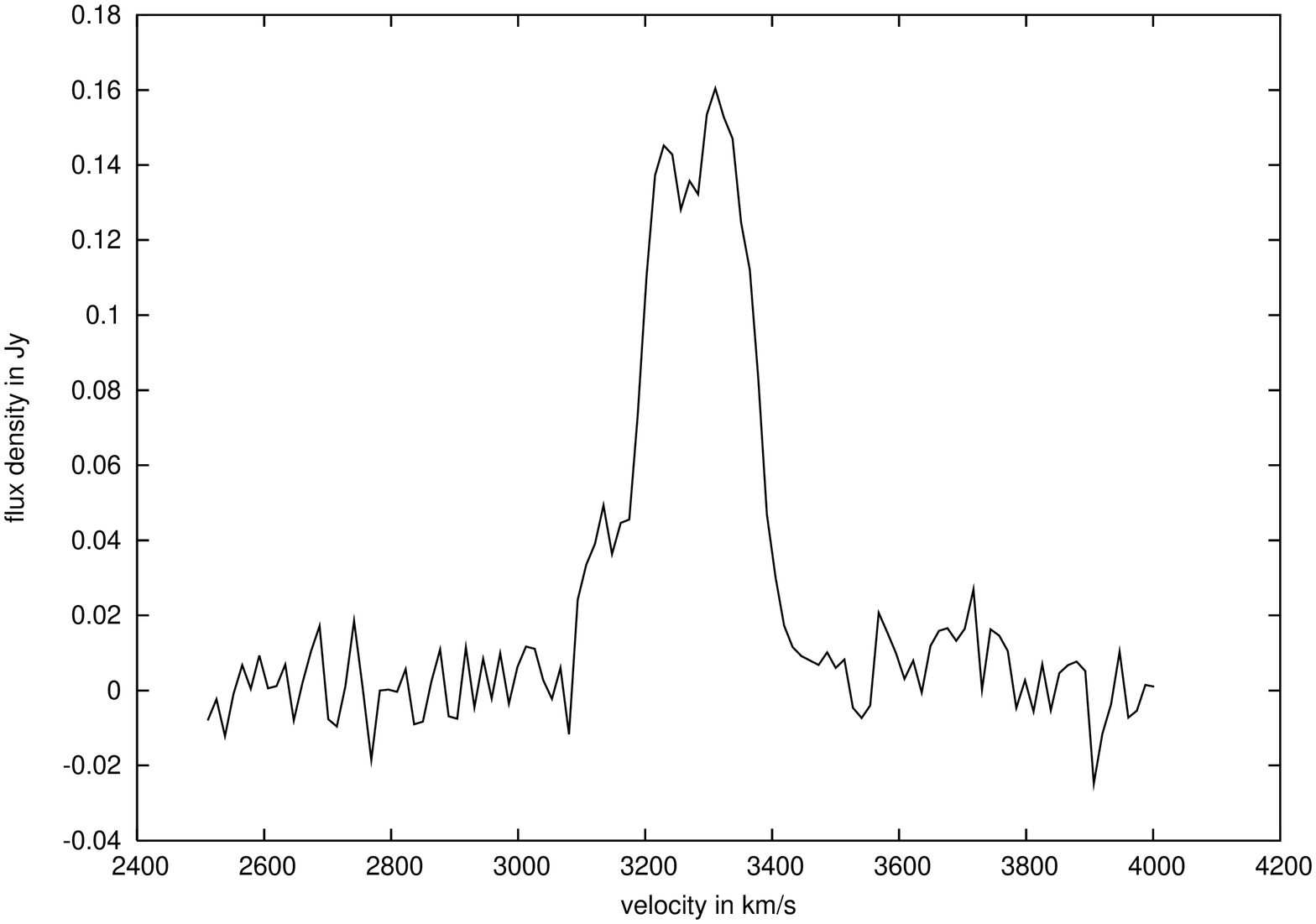}} }
          } 
    \caption{Upper panel: H{\sc i}  column density contours obtained from the low-resolution
      ($\sim$40 arcsec) map, overlaid on the optical  r-band SDSS  image. The H{\sc i}
      column density levels are 1.14$\times$(3, 7, 10, 15, 20, 35, 55, 70, 100, 130, 150, 170)$\times$10$^{19}$ 
      atoms cm$^{-2}$. Beam plotted at bottom left.  Lower panel: H{\sc i} spectrum of the Arp72 system obtained from the 
      low-resolution data cube.
            }
    \label{}
    \end{figure}

A moderate-resolution ($\sim$25 arcsec which corresponds to $\sim$5.3 kpc) image of the Arp72 system is presented
in Fig. 3 overlaid on the far-ultraviolet {\it GALEX} image,  which has an effective 
bandpass of 1350$-$1705 \AA.
The main features are similar to those seen in the low-resolution map, although the ridge of high H{\sc i} 
surface brightness is better defined, and the eastern tail seen in the {\it GALEX} image lies along the ridge line of
the eastern H{\sc i} tail. In the high-resolution image, which has a resolution of $\sim$10 arcsec,
which corresponds to $\sim$2 kpc, only the high H{\sc i} column density regions are seen. Overlays 
of the high-resolution H{\sc i} image of Arp72 system on the 
Spitzer 24-micron image, as well as on the {\it GALEX} UV image are shown in Fig. 4. The 24-micron 
image reveals the relatively new star forming zones and H{\sc i} is seen to be well correlated with the 
bright regions 
in the {\it Spitzer} image. The central star-forming region of the galaxy NGC5996 is devoid of H{\sc i}, 
which is often seen
in many star-forming galaxies, possibly due to a combination of the gas being in molecular form as well 
as ionization 
of the H{\sc i} gas by the hot, young stars. The remaining parts of 
the main disk, including the northern spiral arm, as well as the tidal bridge between the two galaxies show the 
presence of dense H{\sc i} clumps with column densities as high as $\sim$2$\times10^{21}$ atoms cm$^{-2}$.

\begin{figure}
    \hspace{.1cm}
    \centerline{\rotatebox{0}{\includegraphics*[height=3.0in]{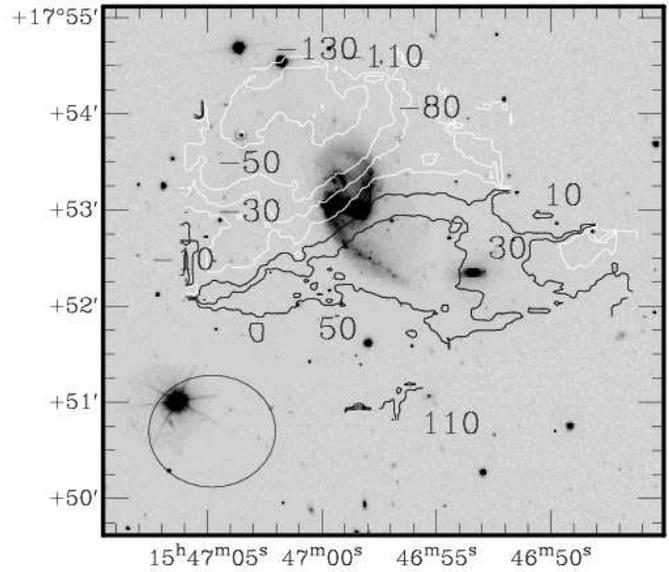}} }
    \vspace*{-0.15in}
    \caption{Radial velocity subtracted H{\sc i} velocity field at low
      resolution ($\sim$40 arcsec overlaid on the  r-band SDSS optical image. The velocities
      plotted are  $-$130, $-$110,
      $-$80, $-$50, $-$30, $-$10, 10, 30, 60, 110 km s$^{-1}$. The negative velocities are in white and the positive velocities are in black contours. Beam plotted at bottom left.
             }     
    \label{}
    \end{figure}
\begin{figure}
    \hspace{.1cm}
    \centerline{\rotatebox{0}{\includegraphics*[height=4.0in]{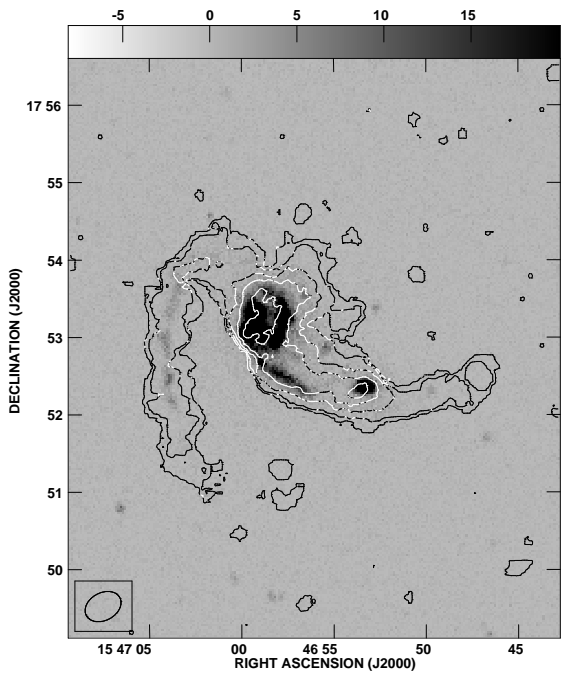}} }
    \vspace*{-0.05in}
    \caption{H{\sc i} column density contours obtained from the medium
      resolution ($\sim$25 arcsec) map, overlaid on the 
      {\it GALEX}  image. 
      The H{\sc i} column density levels are   2.6$\times$(3, 7, 15, 25, 40, 60)$\times$10$^{19}$ atoms cm$^{-2}$. 
            }     
    \label{}
    \end{figure}

\subsection {Radio continuum emission}
The radio continuum image at 1403 MHz with a resolution of $\sim$3 arcsec ($\sim$0.6 kpc) obtained from 
the {\it GMRT} data, is shown in Fig. 5.
The {\it GMRT} and FIRST (Faint Images of the Radio Sky at Twenty Centimeters; Becker, White \& Helfand 1995) 
images are very similar, which show a compact radio source associated with the nuclear region of NGC5996,
along with more diffuse emission, similar to the high-resolution structure seen in many spiral galaxies
(e.g. Saikia et al. 1994). No emission has been detected from the smaller companion NGC5994. The 
NRAO VLA Sky Survey (NVSS; Condon et al. 1998) image also shows no emission from NGC5994, but shows 
an extended tail of emission along a PA of $\sim$22$^\circ$, which extends from NGC5996 well beyond the bridge
of emission connecting the two galaxies. Possible confusion from a background source cannot  be ruled out.
The total flux density of NGC5996 estimated from the {\it GMRT} image is $\sim$22 mJy, 
while the peak and total flux densities, estimated from the NVSS image, including emission from the tail, 
are 20.4 mJy beam$^{-1}$ and $\sim$35 mJy respectively. 
NED lists the continuum flux densities of NGC5996 as 29.6, 32$\pm$4 and $<$30 mJy at 1400, 2380 and 5000 MHz
respectively, taking the values from Condon, Cotton \& Broderick (2002), Dressel \& Condon (1978) and Kojoian et al. (1980)
respectively. Although these values suggest a rather flat radio spectrum at cm wavelengths, suggesting that thermal free-free
emission as well as absorption by dense gas could be playing a significant role, the spectrum needs to be determined 
reliably using measurements at a larger number of frequencies. NGC5996 has been
classified as a starburst, and it would be useful to determine its structure with sub-arcsec resolution to identify
any possible AGN. It is relevant to note that the dominant galaxy in the Arp86 system, NGC7753, which has a
high star-formation rate of $\sim$9 M$_\odot$ yr$^{-1}$, exhibits a flat radio spectrum with $\alpha\sim$0.25 
(S$\propto\nu^{-\alpha}$) between 606 and 1394 MHz \citep{Sengupta09}. In the case of M51, although the integrated spectrum
is steep between 1415 and 2280 MHz with $\alpha\sim$0.87$\pm$0.04, the spectrum of the nuclear region is
relatively flat with $\alpha\sim$0.68$\pm$0.01 (Klein, Wielebinski \& Beck 1984). Studies of the 
central region of M51 show evidence of intense star formation possibly triggered by the interaction. For example, in
the circumnuclear region, Nikola et al. (2001) use PDR models to estimate that the far-ultraviolet field intensities
are similar to those found near OB star-forming molecular clouds in the Milky Way, a few hundred times the local 
Galactic interstellar radiation field. 

\begin{figure*}
    \hspace{.1cm}
    \centerline{\rotatebox{0}{\includegraphics*[width=6.0in]{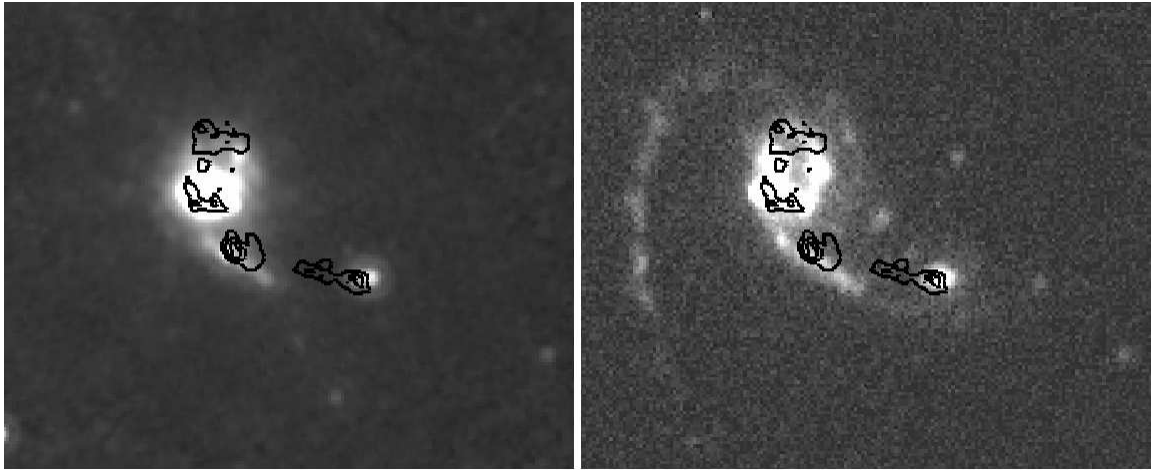}} }
    \caption{H{\sc i} column density contours obtained from the high-resolution
      ($\sim$10 arcsec) map, overlaid on the {\it Spitzer} 24-micron (left panel) and {\it GALEX} far-ultraviolet (right panel)
      images. The H{\sc i} column density levels are (1.7, 2.1)$\times$10$^{21}$ atoms cm$^{-2}$.
            }     
    \label{}
    \end{figure*}

\section {Discussion}

 Sensitive H{\sc i} observations (column density $\le$ 10$^{20}$ atoms cm$^{-2}$) of nearby galaxies 
reveal details which could be related to the galaxy formation and evolution processes.  In addition, to get a 
comprehensive picture of aspects such as the kinematics of the outer disks and warps, star formation 
law at low H{\sc i} column densities, accretion of gas and signatures of dwarf companions around the 
galaxies, deep H{\sc i} imaging of the systems are necessary. In this Section, we discuss our results
on Arp 72, the M51-type system in our sample. 

\subsection{H{\sc i} and star formation in the Arp72 system.} 

The star-formation rate (SFR) in the disks of spirals and starburst galaxies  commonly follows
the relation $\Sigma_{\rm SFR}\propto\Sigma_{\rm gas}^{N}$ with $N\sim$1.4, and is often referred 
to as the Schmidt law
(Schmidt 1959; Kennicutt 1989). The most recent and extensive study of the SFR and star-formation thresholds
in nearby galaxies has perhaps come from the THINGS survey (Leroy et al. 2008; Bigiel et al. 2008).  
Bigiel et al. (2008) find that a molecular Schmidt law with an index $N$=1.0$\pm$0.2 relates $\Sigma_{\rm H2}$
to $\Sigma_{\rm SFR}$ in their sample of spirals. Bigiel et al. (2008) do
not observe a universal relationship between total gas surface density and $\Sigma_{\rm SFR}$ with variations both
within and amongst galaxies. In some cases they find a wide range of SFRs with very similar gas densities, 
while others
seem to follow the Schmidt law over several orders of magnitude in gas surface density. They find the value of $N$ to
vary over a wide range from 1.1 to 2.7, suggesting that there might not be a universal Schmidt law. They also suggest
a link between the environment and relationships between gas and star formation, and star formation is not merely a
function of gas surface density but also the physical conditions that set the ratio of H{\sc i} to H$_2$. Earlier observations also showed evidence of star
formation happening in a wide range of densities and environments. For example, the H{\sc i} bridge in the Magellanic Clouds, 
which has typical column densities $\sim$10$^{20}$$-$10$^{21}$ atoms cm$^{-2}$ is 
known to have star formation occurring in it \citep{Harris}, and no intense 
star formation is seen in the more diffuse Magellanic stream with N(H{\sc i}) 
$\le$3$-$5 $\times10^{20}$ atoms cm$^{-2}$ \citep{Bruns}. 
However, H{\sc ii} regions have been found in tidal debris which are free of any significant H{\sc i} 
association \citep{Ryan-Weber}. 

Our observations of Arp72 also seem to confirm the above findings that there might not be a universal 
Schmidt law. Arp72 is undergoing a stage of intense star formation, possibly triggered by the interaction. NGC5996
has been classified by Balzano (1983) to have a starburst galactic nucleus, and it also shows evidence of
star formation in the disk and spiral arms. The 24-micron band of {\it Spitzer} seems to be 
well suited for tracing recent star formation (Calzetti et al. 2005, 2007).
 \cite{Calzetti} reported a tight correlation between Pa$\alpha$ and 24-micron flux density
in their study of star formation in NGC 5194 (M51a). 
The H$\alpha$ and UV emission have been used traditionally for studying star-formation,  
but both could be significantly affected by dust attenuation, although H$\alpha$ to a much smaller
degree than UV. 
Compared with, say H$\alpha$ which traces early to mid-O type stars with ages less than about
10 Myr, the UV traces somewhat older and less massive stars of O to early-B type with ages
of less than $\sim$400 Myr, providing a star-formation measure over a longer time scale (see 
Calzetti et al. 2007; Smith et al. 2010).  

To investigate the SF region to H{\sc i} column density relation in this system, high resolution H{\sc i} images were overlaid on the {\it GALEX} FUV and Spitzer 24-micron images and compared. A range of H{\sc i} column densities from low to high are seen to be associated with the star forming regions. The high-resolution H{\sc i} image with an angular resolution of $\sim$10 arcsec
($\sim$2 kpc) shows H{\sc i} clumps with column densities as high as $\sim$2$\times10^{21}$ atoms cm$^{-2}$ 
close to the 24-micron peaks of emission in the bridge (Fig. 4). Similar features are seen in the vicinity of the 
companion galaxy NGC5994, and from the disk and towards the north of the main galaxy NGC5996. A prominent tidal tail of H{\sc i} seen towards the east of NGC5996 in the moderate- and low-resolution 
images is not visible in the high-resolution H{\sc i} one (Figs. 3 and 4). This tail towards the east is 
barely visible in the
{\it Spitzer} and SDSS images, but is very clearly seen in the {\it GALEX} image at UV
wavelengths. The H{\sc i} column densities estimated in the eastern tail are in the range of 
0.8 to 1.8$\times$10$^{20}$ atoms cm$^{-2}$, and overlaps with the star-forming region seen most clearly 
in the {\it GALEX} image.  Inspection of this system does not  reveal any prominent star forming region
associated with low ($<$10$^{20}$ atoms cm$^{-2}$) H{\sc i} column densities. 

Using the empirical formula from \cite{CalzettiA} and the 24-micron {\it Spitzer} flux densities from \cite{Smith}, 
the SFR in NGC5996, NGC5994 and the tidal bridge were estimated to be 1.43, 0.06 and 0.21  
M$_{\odot}$ yr$^{-1}$ respectively.  NGC5994 is relatively quiescent compared to NGC5996. 
The high-resolution {\it GMRT} continuum image shows a compact component in the centre of 
NGC5996, and more diffuse emission towards the east and the north, while 
no radio continuum emission has been detected from NGC5994.   
The radio 1400-MHz and far-infrared (FIR) 60 $\mu$m luminosities of NGC5996 are 21.8 and 9.9 
(in log scale) respectively. 
These values are consistent with the radio-FIR correlation \citep{yun}. 

\begin{figure}
    \hspace{.1cm}
    \centerline{\rotatebox{-90}{\includegraphics*[height=4.0in]{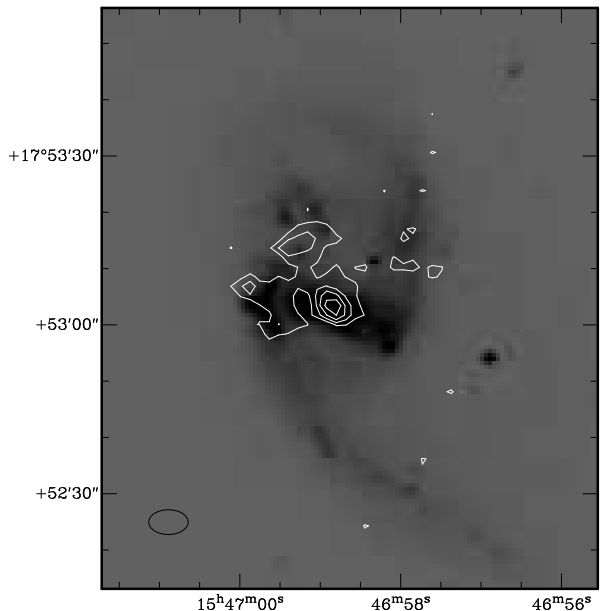}} }
    \vspace*{-0.10in}
    \caption{The {\it GMRT}, 1403-MHz radio continuum image with a resolution of $\sim$2.5 arcsec on optical SDSS. 
             The contour levels are 0.22$\times$(3, 5, 7, 10) mJy beam$^{-1}$. 
            }     
    \label{}
    \end{figure}
\begin{figure}
    \hspace{.1cm}
    \centerline{\rotatebox{-90}{\includegraphics*[height=2.7in]{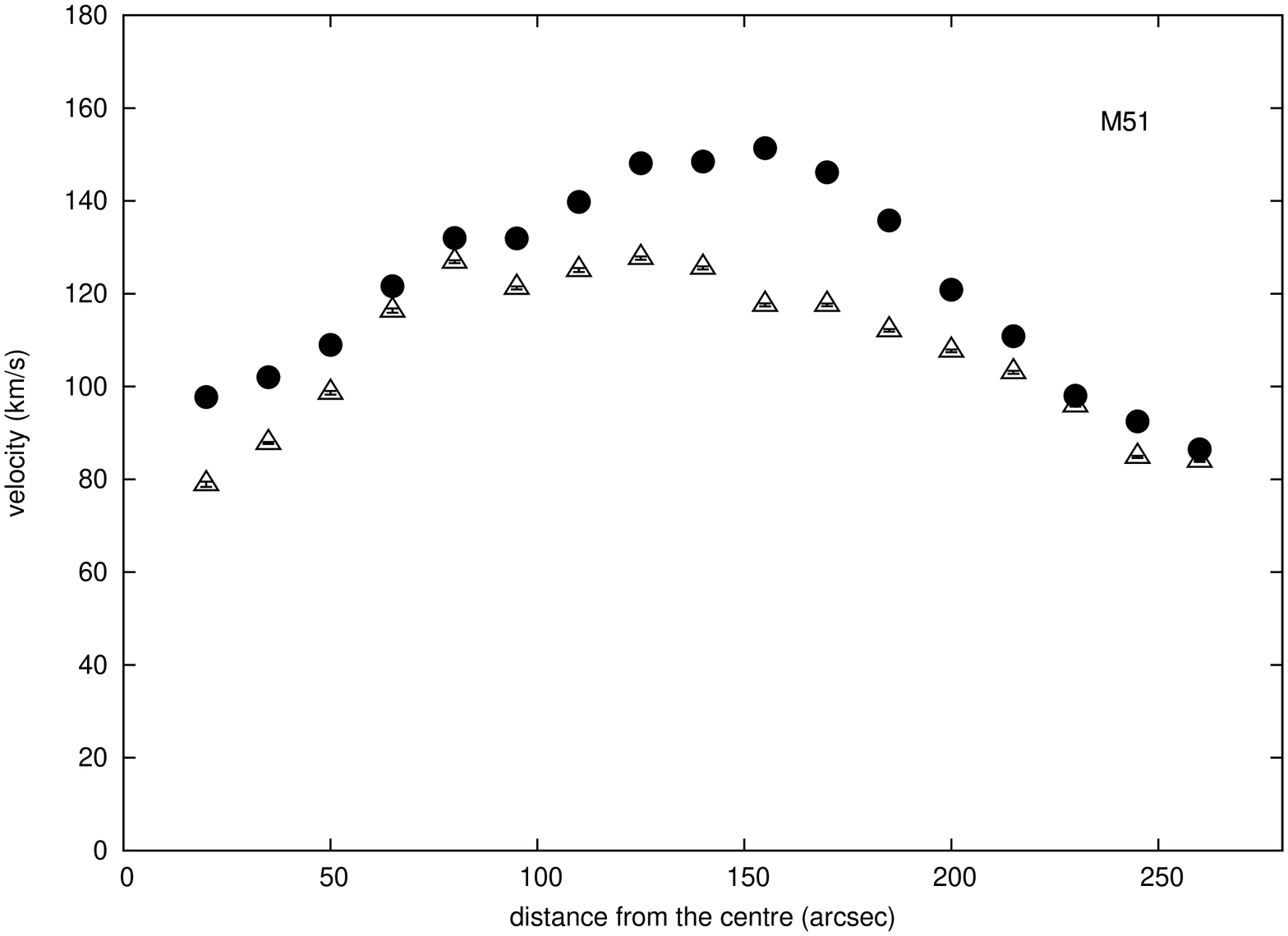}} }
    \centerline{\rotatebox{-90}{\includegraphics*[height=2.7in]{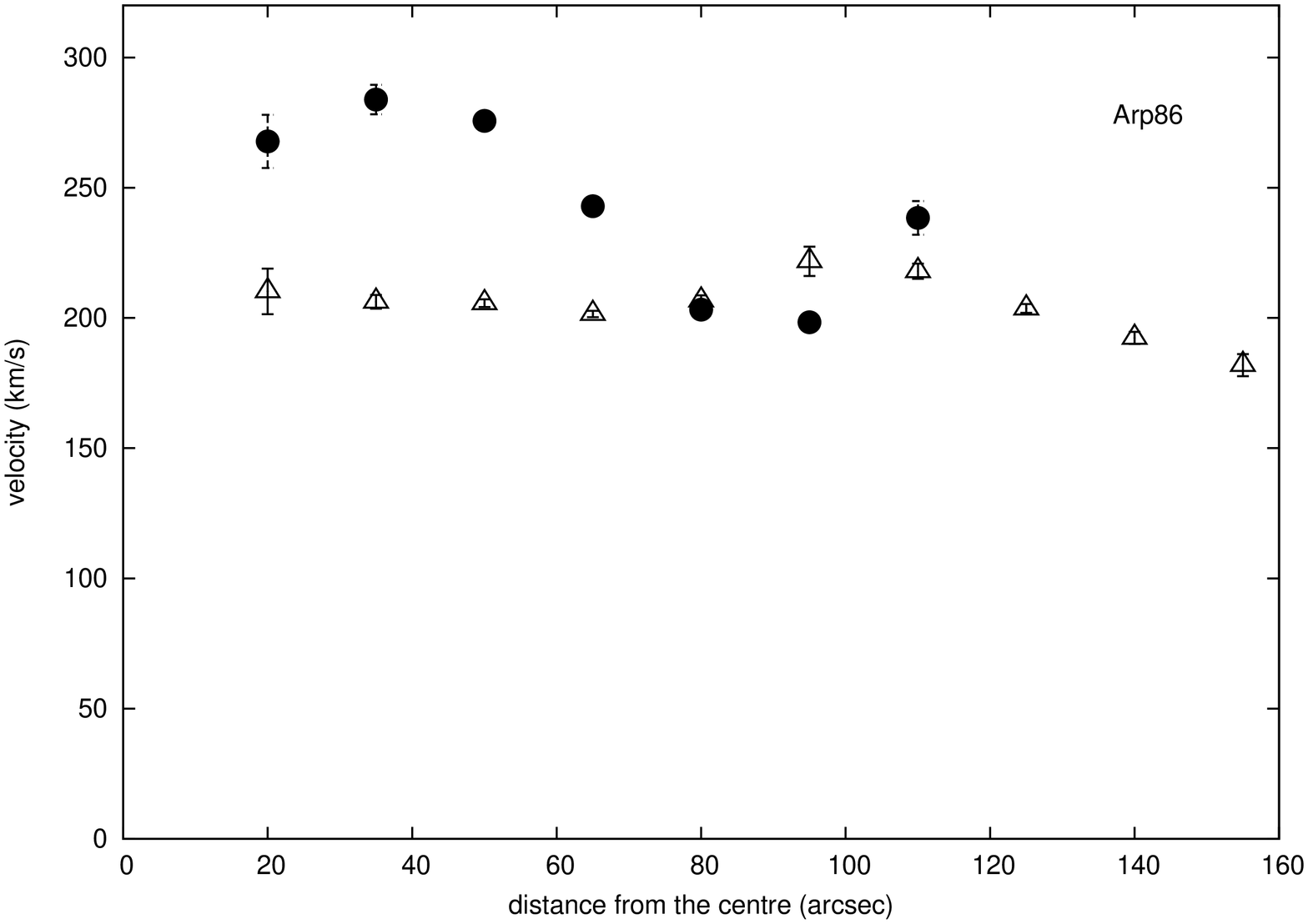}} }
    \centerline{\rotatebox{-90}{\includegraphics*[height=2.7in]{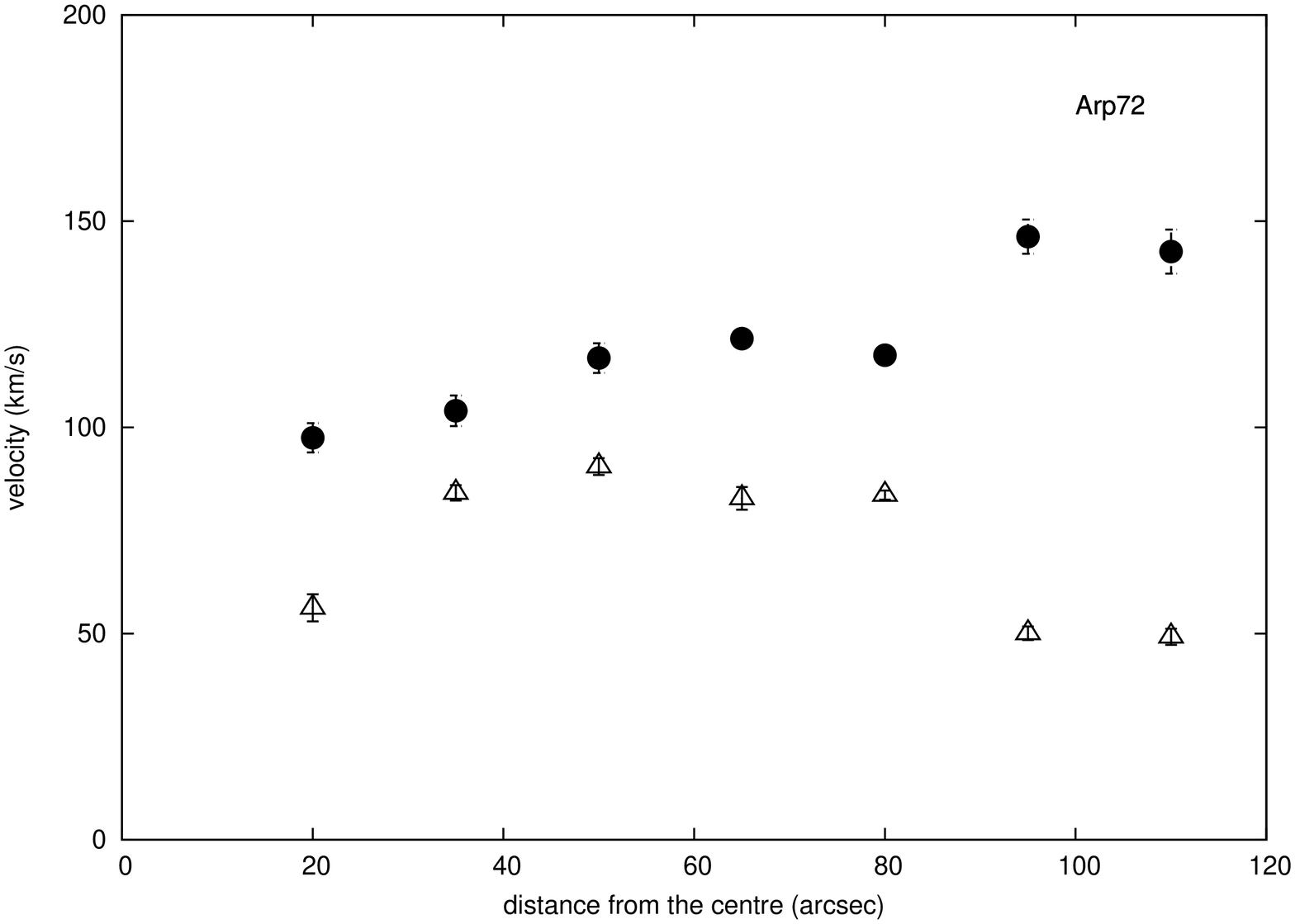}} }
    \caption{ Rotation curves of the approaching and receding sides of the NGC5194 (M51) in the M51 system
              (upper panel), NGC7753 in the Arp86 system (middle panel) and NGC5996 in the Arp72 system
              (lower panel). The triangles indicate the receding side and the filled circles indicate the 
              approaching side.
            }     
    \label{}
    \end{figure}

\subsection {H{\sc i} in M51-type systems}

 As mentioned earlier, the H{\sc i} distribution of Arp72 has been found to be very similar to 
those of M51 and Arp86, categorised as M51-type systems. Although detailed modeling of Arp72 is 
beyond the scope of this paper, we summarise
the similarities in the H{\sc i} morphology and kinematics in these systems, and discuss the
possible kinematic models for such systems (e.g. Toomre \& Toomre 1972; Salo \& Laurikainen 1993;
Theis \& Spinneker 2003).  

\subsubsection {H{\sc i} distribution}
The H{\sc i} observations of M51 \citep{Rots} show that the inner regions
trace the spiral structure as outlined by the star-forming regions, while in the outer regions
of the galaxy faint H{\sc i} extends to larger radii. In addition, there is a prominent curved
H{\sc i} tidal tail which originates at the southern edge of the M51 disk, takes a turn northward 
and extends for $\sim$50 kpc. This feature has no optical counterpart.  
Towards the north, beyond the companion galaxy, a smaller H{\sc i} extension is seen, but it does
not correspond to the faint optical features found near it.
The centre of the main galaxy was seen to be relatively H{\sc i} deficient at high resolution 
\citep{Rots}. 

The H{\sc i} distributions of the M51-type systems we have observed so far, Arp72 and Arp86, have
striking similarities with those of M51. In both cases the centres of the main disks show H{\sc i} 
depletion, and a prominent  H{\sc i} bridge is seen to connect the two 
interacting galaxies. The low-resolution images show that from the edges of the main disks of both 
the systems, a long H{\sc i} tail originates and just like the M51 system, curves back and runs 
almost parallel to the bridge between the two galaxies. Like M51 this extension is seen only in 
lower H{\sc i} column densities. The  H{\sc i} column density limit for this in all the three 
cases is $\le$3$\times$10$^{20}$ atoms cm$^{-2}$. On the opposite side of the disk, towards the 
companion, in all three cases, a similar but smaller low-density H{\sc i} extension without any 
optical counterpart has been seen. 

This apparent similarity in their  H{\sc i} morphology may have its origin in the kind of interactions 
these systems may have gone through. Early test particle simulations by \cite{toomreand} suggested that
the M51 system was experiencing its first close passage encounter.   
However, to explain the new H{\sc i} tidal feature observed by \cite{Rots}, 
about 2 to 3 times longer duration of perturbation was necessary than proposed by \cite{toomreand}, suggesting 
multiple passage encounters \citep{Hernquist}. \cite{Howard} carried out a detailed simulation of the M51 system 
including disk self gravitation and using a three component (stars, gas clouds, inert halo) model. With a single, 
recent passage of the companion past an initially unperturbed disk model, they could explain most of the spiral 
features of the system, but the extended tidal features like the long tail found by \cite{Rots} 
remained unexplained. 
They concluded that  the extended tidal arm seen by \cite{Rots} was in fact a remnant of a previous passage of the 
companion. 

M51 is not an isolated case and two of the M51-type systems we have observed, show similar features and would
thus require the multiple passage model to explain the observed features. As discussed in \cite{Sengupta09}, we 
see that the ratio of the masses of the companion NGC5195 to the main galaxy NGC5194 in the case of the M51 
system, needed to be roughly 0.1 for the simulation to be able to reproduce the observational features \citep{Howard}. 
From the H$\alpha$ rotation curves, the masses of NGC7753 and NGC7752 in the Arp86 system were estimated to be 
1.3$\times$10$^{11}$M$_{\odot}$ and 1.8$\times$10$^{10}$M$_{\odot}$ respectively \citep{marcelin}. This makes the 
mass ratio of the companion to the main galaxy $\sim$0.1, a value similar to that of the 
M51 system \citep{Howard}. For the Arp72 system, the K-band mass ratios for the two galaxies were estimated. 
The estimated mass of NGC5996 and NGC5994 are 2.7$\times$10$^{10}$M$_{\odot}$ and 2.3 $\times$10$^{9}$M$_{\odot}$ 
respectively. The companion to main galaxy mass ratio is again $\sim$0.1, similar to those of the Arp86 and M51 
systems. Thus the mass ratios being also similar in all three systems, is consistent with their interactions being
similar. 

The results from the simulations and rotation curve analysis of the Arp86 system by \cite{Salo} suggests an 
M51-like interaction involving multiple passages of the companion.  
In our study of Arp86, we suggested that the long northern tidal tail may 
be a remnant from the past passage of the companion \citep{Sengupta09}. We notice a similar long tidal arm to the 
north east of the Arp72 system. Though no simulation results exist for this system, from a comparison of the H{\sc i} 
features with those of the other two systems, we suggest that Arp72 has also probably undergone a 
similar multiple passage encounter with its companion. 

\subsubsection{H{\sc i} kinematics}
The velocity field and rotation curves in interacting systems are often found to be disturbed and irregular, 
due to perturbation by the companion galaxy \citep{rubin1983,chengalur1994,Salo2000,fuentes2004}.
Normally, kinematic disturbances are expected to fade out within a few rotation cycles ($\le$ 1 Gyr) \citep{Dale}. 
Therefore these anomalies are of importance as they can be used to trace the interaction history of the 
system \citep{kron}.
One such noticeable signature of perturbation is when the rotation curve of one side of the disk declines while 
the other side remains steady, the so called `bifurcation' of the rotation curve. This happens mainly due to the 
presence of a close companion. Based on such signatures of disturbances, efforts are being made to use them as 
timers of the stage of interaction and understand the possible orbits of the companion \citep{pedrosa, fuentes2004}. 

\cite{Salo2000} investigated the single and multiple passage scenarios for the M51 system,  
and found that high velocity particles in the streamers and also evidence of significant out of plane velocities, support 
a recent perturbation and hence the multiple passage model of interaction.
A comparison of the rotation curves of the single passage and multiple passage encounter simulations showed that 
in a single passage the curve remains similar to the circular velocity curve and in the multiple passage the curve 
declines after a certain radius. 
Although H{\sc i} images and velocity fields of M51 and Arp86 have been published earlier \citep{Rots, Sengupta09}, 
the H{\sc i} rotation curves for the dominant galaxies of the M51, Arp86 and Arp72 systems are presented for the first time in Fig. 6. 
The companion galaxies are too small to obtain rotation curves. THINGS data have been used 
to obtain the rotation curve of NGC5194 (M51), while  the rotation curves of NGC7753 (Arp86)  and NGC5996 (Arp72) 
have been derived using the {\it GMRT} data. The  Groningen Image Processing System ({\tt GIPSY}), has been used to obtain 
the rotation curves from the FITS files of the velocity fields. 
The best fit curves for all the three systems have been obtained by keeping the inclination angle and centre coordinates 
as fixed parameters and rotational velocity and  position angle as free parameters in the {\tt `ROTCUR'} program of 
{\tt GIPSY}. 

For NGC5194, both the curves show signs of declining velocity towards the edge, the approaching side falling faster 
than the receding side (Fig. 6, upper panel).  The companion is at the tip of the approaching side, about 300 arcsec away.  The `S'-shaped profile  is also 
seen for the H${\alpha}$ rotation curves observed for the M51 system \citep{Tilanus91}. 
A similar trend of declining  H{\sc i} rotation curve is seen on one side of the Arp86 system, where the companion is 
present. The side nearer to the companion, shows higher rotation velocity and faster decline near $\sim$100 $\arcsec$ 
from the centre (Fig. 6, middle panel), which is roughly the distance to the companion. The curve on the receding side 
remains undisturbed 
and flat till a longer distance of $\sim$160 $\arcsec$. The decline of rotation velocity near 100$\arcsec$  is in agreement 
with the results of \citep{marcelin}, where the authors report a falling  H${\alpha}$ rotation curve for the western 
side of NGC7753.  The rotation curves of NGC5996 of Arp72 are shown in Fig. 6 (lower panel). 
A declining  H{\sc i} rotation curve for the receding side of NGC5996 is again apparent. The curve shows signs of a rapid decline 
near a radial distance of $\sim$80$\arcsec$ from the centre, which is roughly the distance to the companion. It would be interesting to examine the H{\sc i} rotation curves with data at other wavelengths.
Thus for all these systems, we do notice a similar nature of declining rotation curves, especially on the side where the 
companion is present, suggesting a similarity with the results of \cite{Salo} that a multiple passage model can 
often explain these systems better than a single passage model as proposed by  \cite{toomreand}. 
    
\section{Concluding remarks}
{\it GMRT} observations  of the interacting galaxies NGC5996 and NGC5994, which make up the Arp72 system 
show a complex distribution of H{\sc i} tails and a bridge between the two galaxies due to tidal interactions 
between them. Optical, infrared and ultraviolet observations all show a bridge between the two galaxies, and
a tail on the eastern side, which is seen most clearly in the {\it GALEX} image at ultraviolet wavelengths. A range of H{\sc i}  column densities from $\sim$1.7$-$2$\times$10$^{21}$ atoms cm$^{-2}$ to  0.8$-$1.8$\times$10$^{20}$ atoms cm$^{-2}$  are seen to be associated with sites of star formation in the bridge, disk of 
the main galaxy and the tidal debris. There is a depletion of H{\sc i} in the centre of the more
massive galaxy, possibly due to the gas being in the molecular phase as well as ionization of the gas by
the starburst in the central region of NGC5996. No star forming zone has been found to be completely free of any H{\sc i} association or correlated with very low H{\sc i} column densities.      
The morphological and kinematic similarities of Arp72 with M51 and Arp86, suggest 
a multiple passage model of \cite{Salo} is preferred over the single passage model of \cite{toomreand}, to understand the H{\sc i} features in M-51 systems. 
 The striking similarities for the three M51-type systems discussed here suggest similar time-scales.
Howard \& Byrd (1990) estimate the period of the companion's orbit to be 500 Myr, and that it
merges in less than three orbits. The orbital period for Arp86 is similar to within a factor of $\sim$2
(see Salo \& Laurikainen 1993). Identification of such systems at high redshifts from sensitive H{\sc i}
surveys using the upcoming telescopes and modelling the kinematics would provide valuable inputs towards
estimating the range of time scales of minor mergers and our understanding of galaxy evolution. 
A detailed study of Arp72 at H$\alpha$ and say, CO, to compare the kinematics of the ionized 
and molecular gas with the H{\sc i} gas would provide a more complete picture of the kinematics and star 
formation of this interesting M51-type system.  

\section{Acknowledgments}
We thank an anonymous reviewer for his/her comments which helped improve the manuscript, and 
Beverly Smith for providing us very promptly with the files of the images.
We thank the staff of the {\it GMRT} who have made these observations possible. The {\it GMRT} is operated by 
the National Centre for Radio Astrophysics of the Tata Institute of Fundamental Research. This research 
has made use of the NASA/IPAC Extragalactic Database (NED) which is operated by the Jet Propulsion Laboratory, 
California Institute of Technology, under contract with the National Aeronautics and Space Administration.

\end{document}